\documentclass[aps,prl,preprint,showpacs,preprintnumbers,amsmath,amssymb]{revtex4}

\usepackage{graphicx}
\usepackage{dcolumn}
\usepackage{bm}

\begin{document}

\title{Can a Hagedorn system have a temperature other than $T_C$ \\ 
or \\
can a thermostat have a temperature other than its own?}

\author{L. G. Moretto, K. A. Bugaev, J. B. Elliott and L. Phair}
\affiliation{Nuclear Science Division, Lawrence Berkeley National Laboratory,
Berkeley, CA 94720}

\date{\today}
\begin{abstract}
This is a note intended to complement our paper (nucl-th/0504010) and addressed to the attention of QGP workers interested in  bag models, Hagedorn spectra, and the like. It tries to show that with a Hagedorn-like experimental spectrum the partition function can not be calculated and that a canonical description derived for the microcanonical ensemble exists only for a single, fixed temperature.
\end{abstract}

\preprint{LBNL-59191}
\maketitle

\section{Introduction}

In a recent paper (nucl-th/0504010), we have shown that systems with exponential spectra $\rho\propto\exp (aE)$ such as the Hagedorn mass spectrum and the bag model mass spectrum act as thermostats at temperature $T_C=1/a$ and impart this unique temperature to any system thermally coupled to them. In contrast to this, the standard wisdom asserts that such systems admit any temperature $T<T_C$ according to their ``alleged'' partition function $Z(T)=T_C T/(T_C-T)$. 

In what follows we try to show that the above partition function is incorrect and leads to erroneous results when applied to resonance gases, etc.

\section{An interesting exercise}

Consider a system $A$ composed of ice and water at standard pressure. For such a system the temperature is $T_A=273$ K. Because of coexistence, we can feed or extract heat to/from the system without changing $T_A$. The system $A$ is a thermostat.

If a quantity $Q$ of heat is added to the sytem, the change in entropy is
\begin{equation}
\Delta S = \frac{Q}{T_A}.
\end{equation}
The level density of $A$ is then 
\begin{equation}
\rho(Q) = S_0e^{Q/T_A}\approx Ke^{E/T_A}.
\end{equation}

The level density, or spectrum, is exponential in $E$ and depends only on the intrinsic ``parameter" $T_A$. Let us calculate the partition function of $A$:
\begin{equation}
\label{ }
Z(T) = \int e^{E/T_A} e^{-E/T} dE = \int e^{-\left(\frac{1}{T}-\frac{1}{T_A}\right) E} dE \\
= \frac{T_A T}{T_A-T}
\end{equation}

This seems to indicate that $A$ can assume {\em any} temperature $0\le T<T_A$.

But, by hypothesis, the only temperature possible for $A$ is $T_A$. What is the trouble?

\section{The canonical ensemble and the ``implicit" thermostat}

Let us consider two systems $A, B$ with level densities $\rho _A$ and $\rho _B$. Let the systems be thermally coupled to each other with total energy $E$. We now calculate the distribution in energies between the two systems,
\begin{equation}
\label{ }
\rho _T(x) = \rho_A(E-x)\rho _B(x)
\end{equation}
Let $A$ be a ``thermostat", i.e. $\rho_A=e^{\epsilon /T_A}$. Then 
\begin{equation}
\label{ }
\rho_T(x) = \exp\left(\frac{E-x}{T_A}\right) \rho_B(x) = e^{E/T_A}e^{-x/T_A}\rho_B(x).
\end{equation}
Let us integrate over $x$
\begin{equation}
\label{ }
\int\rho_T(x) dx = e^{E/T_A}\int e^{-x/T_A}\rho_B(x) dx = e^{E/T_A} Z_B(T_A).
\end{equation}
This is the origin of the partition function $Z_B(T_A)$ and the meaning of ``implicit" thermostat. By changing ``thermostat" we can change $T_A$ and the temperature of $B$.

Thus, every time we construct a partition  function, we imply the gedanken experiment of connecting the sytem to a thermostat, and that this experiment is actually possible for the system we are studying.

Does this always work?

To see this, let us look for the most probable value of the distribution $\rho_T(x)$, which defines the equilibrium partition, by taking the log and differentiating:
\begin{equation}
\label{ }
\ln \rho_T(x) = \ln \rho_A(E-x) + \ln \rho_B(x)
\end{equation}

\begin{equation}
\label{ }
\frac{\partial\ln\rho_T(x)}{\partial x} = -\left.\frac{\partial\ln\rho_A}{\partial x}\right| +  \left.\frac{\partial\ln\rho_B}{\partial x}\right| = 0 
\end{equation}
or
\begin{equation}
\label{ }
\frac{1}{T_A} = \frac{1}{T_B}.
\end{equation}
For this to be possible, it is necessary that $\rho_A$ and $\rho_B$ admit the {\em same} logarithmic derivative somewhere in the allowed range of energy $x$.

Usually, and always for concave functions (convex functions are anathema!), $S(x) = \ln\rho(x)$ and $T=(\partial S/\partial x)^{-1}$ is such that  $0\le T\le \infty $.  Thus, for such systems it is possible to match derivatives for whatever value of $E$. Thermal equilibrium is achievable only at the temperature of the thermostat.

However, if $S_A(E) = \ln\rho_A(E)$ is linear in $E$, then $T_A=(\partial S/\partial E)^{-1}$ is a constant, independent of $E$.
In this case, it is up to $B$ to look for the value of $x$ at which its logarithmic derivative matches $1/T_A$. The system $A$ is a ``thermostat" at $T=T_A$ and $B$ can only try to assume the value $T=T_B=T_A$, if it can do so.

Now suppose that also $S_B(E) = \ln\rho_B(E)$ is linear in $E$ with an inverse slope $T_B$. This means that only if $T_A=T_B$ is equilibrium possible, and the partition function of $B$, $Z_B$ is meaningfully defined only for $T=T_B$ and not for $0\le T\le T_B$. We cannot force a temperature $T\ne T_B$ on a thermostat. It can only have its own intrinsic temperature $T_B$.

Placing systems $A$ and $B$ into contact will lead to a continuous heat flow from one system to the other. Thermal equilibrium is not achievable.

To summarize: it is permissible to calculate the partition function of a system only if its $S(E)$ admits as inverse derivatives values such as we want to impose through our Laplace transform. Failing that, the resulting partition function does not satisfy any thermodynamic criterion.


\begin{thebibliography}{99}





\end{thebibliography}
\end{document}